\documentstyle[sprocl]{article}

\input{psfig}
\def\Journal#1#2#3#4{{#1} {\bf #2}, #3 (#4)}

\def\PRB{{\em Phys. Rev.} B}

\def\be{\begin{equation}}
\def\ee{\end{equation}}
\def\bea{\begin{eqnarray}}
\def\eea{\end{eqnarray}}
\begin{document}

\title{INTER-LAYER EDGE TUNNELING AND TRANSPORT PROPERTIES
IN DOUBLE-LAYER QUANTUM HALL SYSTEMS}

\author{D. YOSHIOKA}

\address{Department of Basic Science, University of Tokyo, Komaba,\\ Tokyo 153,
Japan}
%World Scientific Publishing Co, 1060 Main Street,
%River Edge,\\ NJ 07661, USA}

\author{ A.H. MACDONALD }

\address{Department of Physics, Indiana University, Bloomington,\\ IN
47405, USA}


\maketitle\abstracts{
A theory of transport in the quantum Hall regime is developed for
separately contacted double-layer electron systems.
Inter-layer tunneling provides a channel
for equilibration of the distribution functions in the two layers
and influences transport properties through the resulting
influence on steady-state distribution functions.
Resistences for various configurations of the electrodes
are calculated as a function of the inter-layer tunneling amplitude.
The effect of misalignment of the edges of the two layers
and the effect of tilting the magnetic field away from the
normal to the layers
on the inter-layer tunneling amplitude near the sample edges
are investigated.
The results obtained in this work is consistent with recent experiments.
}

\section{Introduction}
%\subsection{Producing the Hard Copy}\label{subsec:prod}

Recently it has become possible to fabricate a multi-layer quantum Hall system,
where current and voltage leads are attached separately
to each layers.\cite{eisen}
For such a system it is expected that inter-layer interaction affects
the resistances.
For example, in a recent experiment by Ohno {\it et al}.,~\cite{ohno}
the Hall resistance and
the longitudinal resistance on one layer is affected considerably
by the presence of the other.
Inspired by this experiment, we have investigated how the interlayer tunneling
affects the resistance in the quantum Hall regime, where the electron
distribution near the system edges determine the
electrical conduction.\cite{yoshi}
In Sect.2  we explain our theory briefly.
We show the results where only one of the layer have leads attached.
It is shown that the resistances are non-local, and depends on the
strength of the inter-layer tunneling probability.
Then in Sect. 3 we give our new result on how misalignment of the sample
edges and tilting of the
magnetic field affects the hopping probability.
Finally brief discussions are given.

\begin{figure}
%\rule{5cm}{0.2mm}\hfill\rule{5cm}{0.2mm}
%\vskip 4.0cm
%\rule{5cm}{0.2mm}\hfill\rule{5cm}{0.2mm}
\psfig{figure=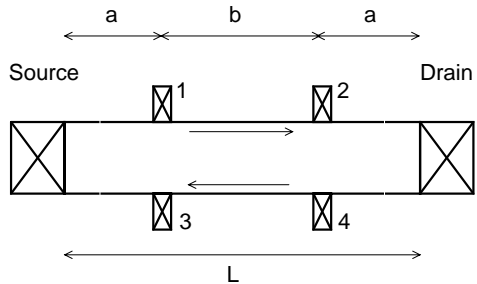,height=1.5in}
\caption{Geometry of the sample.
Two identical layers are stacked vertically.  The leads
(source, drain, 1, 2, 3, and 4) can contact both layers or either layer.
The drift directions along the edges are indicated by arrows.
The lengths $L$, $a$, and $b$ specify the length of the sample and
the positions of the leads.
\label{fig1}}
\end{figure}

\section{Resistance}

We consider a Hall bar type sample shown in Fig.1.
The energy levels in the bulk are aligned in the two layers,
and we consider the system in the lowest quantum Hall regime.
%Namely the Fermi level is between the lowest and the second lowest Landau
%levels, so the lowest Landau levels are filled completely.
For a single layer system the
quantum Hall effect can be understood as a consequence of the spatial
separation between left-going and right-going states on opposite edges  of the
sample.\cite{butt}
At each edge we can define the chemical potential, which stays constant
between the leads, and which differs between the opposite edges.
However, in the case of a double layered system with finite inter-layer
tunneling,
the local chemical potential in each layer is generally not constant
between the leads.
We take $x$-axis along the edge and introduce $\mu_\sigma (x)$ for
the local chemical potential, where $\sigma = \pm$ represents the layer
index.
Then the following equation governs the development of the chemical
potentials between the leads:
\begin{equation}
\frac{d\mu_{\sigma}(x)}{dx} = - \frac{1}{\xi}[\mu_{\sigma}(x)
-\mu_{-\sigma}(x)],
\label{boltchem}
\end{equation}
where $\xi$ is a parameter meaning the relaxation length for interedge
equilibration.
%This equation is easily solved,
%\begin{equation}
%\mu_{\sigma}(x) = \frac{1}{2}(1+{\rm e}^{-2x/\xi})\mu_{\sigma}(0)
%+ \frac{1}{2}(1-{\rm e}^{-2x/\xi})\mu_{-\sigma}(0).
%\label{chemsolv}
%\end{equation}
The current along an edge in layer $\sigma$ at position $x$ is given  by
\begin{equation}
I_{\sigma} (x) = \frac{e}{h}[\mu_{\sigma}(x) - \epsilon_{0}],
\label{currentsolv}
\end{equation}
with $\epsilon_{0}$ being a reference energy.
The voltage leads will not affect the chemical potential, since the
current will not flow through the lead.
%and the chemical potential is observed there.
On the other hand the current leads give discontinuous change
to the chemical potential.
%, the discontinuity being proportional to
%the current extracted or introduced at the lead.
Following these rules we can solve for the chemical potentials of both
layers, once the external current through each leads are given for any
configuration of the leads.

As an example of such a solution, the followings are the obtained resistances
for the case where only the minus layer has leads attached.
Here current is fed to the minus layer through the source and extracted through
the drain,
and $\mu_-$ is observed at voltage leads 1 to 4 (see Fig.1).
The longitudinal resistance $R_{xx} =R_{12} = R_{34}$ and the Hall resistance
$R_{xy} = R_{13}$ are given as follows:
\begin{equation}
R_{xx} = \frac{h}{e^{2}}
\frac{\exp(-2a/\xi)[1 - \exp(-2b/\xi)]}{2[1+\exp(-2L/\xi)]},
\label{rxx}
\end{equation}
\begin{equation}
R_{xy} = \frac{h}{e^{2}}
\frac{1+\exp(-2a/\xi)+\exp[-2(L-a)/\xi] + \exp(-2L/\xi)}
{2[1+\exp(-2L/\xi)]}.
\label{rxy}
\end{equation}
where $a$ and $b$ specify the voltage probe
positions as indicated in Fig.~1.
Thus only in the strong (weak) coupling limit, where $L/\xi = \infty (0)$,
the resistances are quantized: $R_{xx}=0$ and $R_{xy} = h/2e^2 (h/e^2)$.

\section{Relaxation length $\xi$}

The results for the resistances depend on the ratio $L/\xi$.
Thus we need to know typical size of this parameter, and how it depends
on various factors.
The tunneling between the layers occur between state with the same energy.
Let's assume such pair of states are given by wave functions
$\phi_\sigma(x,y,x)$ in the absence of tunneling.
When the two edges of the two layers are aligned perfectly,
they satisfy $\phi_+(x,y,z) = \phi_-(x,y,z-d)$, where $d$ is the interlayer
separation.
In the presence of the tunneling, they couple each other, and
symmetric and antisymmetric combination of these states will
become the eigen states.
{}From the energy difference of these states, $\Delta_{SAS}$,
the correlation length is estimated to be $\xi = \hbar v/\Delta_{SAS}$,
where $v \simeq \ell\omega_c$ is the velocity of edge states,
with $\ell$ being the magnetic length
and $\omega_c$ being the cyclotron frequency.
This estimate gives $L/\xi \simeq 50$ for $L=200\mu$m,
and $\Delta_{SAS}=0.02$meV, which are typical values for an
experimental situation.~\cite{ohno}
This value of $L/\xi$ is quite large, and the system can be considered to
be in the strong coupling regime.
However, various factors reduce this value.
Among them here we consider misalignment of the edges and in-plane magnetic
field.
We assume the center coordinate of $\phi_+$ projected on to the
$xy$-plane cross many times with that of $\phi_-$.
Among those crossings we focus on one of them,
approximate the trajectory of the center coordinates
around the crossing by straight lines,
and calculate the contribution to the tunneling amplitude from a section of
length $L_s (\gg \ell)$ around the crossing.
We take the coordinate such that the $x$-axis bisects the two edges,
so the two edges make angle $\pm \theta/2$ from the $x$-axis.
The in-plane field $B_x$ and $B_y$ also refer to this coordinate.
The tunneling amplitude, $I(B_x,B_y,\theta)$,
is proportional to the overlap integral of
$\phi_+$ and $\phi_-$.
If we approximate $\phi_\sigma$ by the lowest Landau level wave function,
the relative amplitude at $\theta=0$ is given as follows:
\be
{{I(B_x,B_y,0)}\over{I(0,0,0)}} =
 {{2\ell}\over{L_s}}~{{\sin(L_s\beta_y/2\ell)}\over{\beta_y}}
\exp(-{{\beta_x^2}\over{4}}),
\label{zeroheta}
\ee
where $\beta_x = B_xd/B_z\ell$, and $\beta_y = B_yd/B_z\ell$.
So in this case, $B_y$ has a drastic effect: the amplitude is deminished by
a small $B_y$, but $B_x$ dependence is small.
For $\theta \gg (\ell/L_s)^2$ the amplitude is much reduced and given by
\be
{{I(B_x,B_y,\theta)}\over{I(0,0,0)}}
 = {{\ell}\over{L_s}}~\sqrt{{{2\pi}\over{\sin\theta}}}
\exp[-{{(\beta_x^2+\beta_y^2)}\over{4}}].
\label{theta}
\ee
In this case $I(B_x,B_y,\theta) \ll I(0,0,0)$,
and the effect of the in-plane magnetic field is quite weak.

\section{Discussion}
In recent experiments Ohno {\it et al.}~\cite{ohno}
observed $R_{xy} \simeq h/e^2$ in the
presence of finite $R_{xx}$ in the situation considered in Sec.2.
This result is at a first glance inconsistent with our result, since
$L/\xi \simeq 50$, if estimated using $\Delta_{SAS} = 0.02$meV.
However, the inconsistency is resolved, if the edges are not aligned
perfectly.
In this case the tunneling amplitude is reduced as shown in Eq. 6.
In the experiments they observed that
the effect of in-plane field is quite small.
This insensitivity to the tilting is also in accordance with Eq. 6.
Thus our theory is consistent with the experiments.
It is general enough to treat various other configurations of
leads.
We hope such situations are realized experimentally, and results
are compared with our theory.

\end{document}